\documentclass[conference]{IEEEtran}
\IEEEoverridecommandlockouts

\usepackage{amsmath,amssymb,amsfonts}
\usepackage{bm}
\usepackage{algorithmic}
\usepackage{graphicx}
\usepackage{textcomp}
\usepackage{xcolor}
\usepackage{array}
\usepackage{booktabs}

\makeatletter
\newcommand{\linebreakand}{%
  \end{@IEEEauthorhalign}
  \hfill\mbox{}\par
  \mbox{}\hfill\begin{@IEEEauthorhalign}
}
\makeatother

\begin{document}

\title{Cross-attention Inspired Selective State Space Models for Target Sound Extraction
\thanks{This work is supported by the National Natural Science Foundation of China (No.61175043, No.61421062), and the Highperformance Computing Platform of Peking University.}
}

\author{\IEEEauthorblockN{1\textsuperscript{st} Donghang Wu}
\IEEEauthorblockA{
\textit{National Key Laboratory of General Artificial Intelligence}\\
\textit{School of Intelligence Science and Technology} \\
\textit{Peking University}, Beijing, China \\
sdlwwdh@pku.edu.cn}
\and
\IEEEauthorblockN{2\textsuperscript{nd} Yiwen Wang}
\IEEEauthorblockA{\textit{National Key Laboratory of General Artificial Intelligence}\\
\textit{School of Intelligence Science and Technology} \\
\textit{Peking University}, Beijing, China \\
pku\_wyw@pku.edu.cn}
\linebreakand 
\IEEEauthorblockN{3\textsuperscript{rd} Xihong Wu}
\IEEEauthorblockA{\textit{National Key Laboratory of General Artificial Intelligence}\\
\textit{School of Intelligence Science and Technology} \\
\textit{Peking University}, Beijing, China \\
wxh@cis.pku.edu.cn}
\and
\IEEEauthorblockN{4\textsuperscript{th} Tianshu Qu}
\IEEEauthorblockA{\textit{National Key Laboratory of General Artificial Intelligence}\\
\textit{School of Intelligence Science and Technology} \\
\textit{Peking University}, Beijing, China \\
qutianshu@pku.edu.cn}
}

\maketitle

\begin{abstract}
The Transformer model, particularly its cross-attention module, is widely used for feature fusion in target sound extraction which extracts the signal of interest based on given clues. Despite its effectiveness, this approach suffers from low computational efficiency. Recent advancements in state space models, notably the latest work Mamba, have shown comparable performance to Transformer-based methods while significantly reducing computational complexity in various tasks. However, Mamba's applicability in target sound extraction is limited due to its inability to capture dependencies between different sequences as the cross-attention does. In this paper, we propose CrossMamba for target sound extraction, which leverages the hidden attention mechanism of Mamba to compute dependencies between the given clues and the audio mixture. The calculation of Mamba can be divided to the query, key and value. We utilize the clue to generate the query and the audio mixture to derive the key and value, adhering to the principle of the cross-attention mechanism in Transformers. Experimental results from two representative target sound extraction methods validate the efficacy of the proposed CrossMamba.
\end{abstract}


\section{Introduction}
Target sound extraction (TSE) aims to separate the sound of interest from a signal mixture using given clues, such as the target sound class \cite{waveformer}\cite{soundbeam}, the pitch information \cite{wywpitch}, an enrollment audio sample \cite{spex}\cite{speakerbeam}\cite{enrolltse1} or visual stimuli like lip movements \cite{avsepformer}\cite{tdfnet}\cite{mu2024separate}. The typical architecture of TSE models comprises an audio encoder and a clue encoder, which independently encode the audio mixture and the clues into their respective feature representations. Additionally, a separator module performs feature fusion of the encoded audio and clues to extract the target components from the audio features.

A crucial component of a target sound extraction model is the feature fusion method, which captures the dependencies between the audio mixture and the provided clues. Various feature fusion techniques have been explored, such as element-wise multiplication \cite{soundbeam}, concatenation \cite{voicefilter}, and Feature-wise Linear Modulation (FiLM) layers \cite{film}. Recently, with the Transformer and its attention mechanism \cite{transformer} setting new performance benchmarks across various fields, including language modeling \cite{devlin2018bert}, image processing \cite{touvron2021training}, and audio signal processing \cite{dptnet}, the Transformer decoder, particularly its cross-attention modules, has been increasingly used for feature fusion in TSE tasks. For instance, Waveformer employs the Transformer decoder with cross-attention to extract the target sound using the class label as the clue \cite{waveformer}. Additionally, the Transformer decoder and its cross-attention module have been utilized in AV-SepFormer to incorporate visual clues for target sound extraction \cite{avsepformer}. Although Transformers are highly effective at capturing long-range dependencies, their computational complexity and memory demand are typically high. This is due to the fact that the memory demands and the computational complexity of the attention mechanism in Transformers increase quadratically with the sequence length.

The structured state space model (S4) has been developed for modeling long sequences. S4 can be viewd as convolutional neural networks (CNN) for parallelizable training and recurent neural networks (RNN) for efficient linear time complexity inference \cite{s4}. Selective state space model (Mamba) further extends S4 by making the parameters input-dependent \cite{mamba}. Mamba has been demonstrated to be a computational efficient alternative to Transformers in natural language \cite{mamba}, image \cite{umamba} and audio signal processing \cite{spatialnet}\cite{mamba_audio}. However, Mamba is designed only to model long-range dependencies within a single sequence and cannot handle interactions between different sequences. This limitation restricts its use in target sound extraction tasks.


In this paper, a cross-attention based Mamba network named CrossMamba is proposed for feature fusion in target sound extraction. Following the analysis in \cite{mamba_atten}, we decompose the calculation of Mamba to the query, key and value. Then the clue is utilized to generate the query and the input signal mixture is leveraged to generate the key and value. Our CrossMamba is evaluated on two representative target sound extraction models: AV-SepFormer \cite{avsepformer} and Waveformer \cite{waveformer}. Experimental results demonstrate that CrossMamba is both resource-efficient and effective in performing target sound extraction using various types of clues.

The rest of this paper is organized as follows. Section \ref{CrossMamba} introduces the preliminary knowledge and the design principles of CrossMamba. Section \ref{CMTSE} details the application of CrossMamba in two representative target sound extraction models: AV-SepFormer and Waveformer. Section \ref{experiments} validates the effectiveness of CrossMamba through experimental results. Finally, our conclusion is presented in section \ref{conclusion}.



\section{CrossMamba}
\label{CrossMamba}
The design of CrossMamba is based on the principle of the cross-attention mechanism and the hidden attention matrix in Mamba's calculation. This section begins by introducing the principles of cross-attention and Mamba, followed by a description of CrossMamba based on these principles. 
\subsection{Attention mechanism}
The input to an attention function consists of the query $\bm{Q} \in R^{N\times d_k}$, key $\bm{K}\in R^{N\times d_k}$ and value $\bm{V}\in R^{N\times d_v}$, which are derived from input sequences. Here, $N$ represents the sequence length, $d_k$ denotes the dimension of the query and key and $d_v$ indicates the dimension of the value. The output of the attention can be viewed as a weighted sum of the value, where the weights are determined by the correlation between the query and key, which can be formulated as
\begin{equation}
\begin{split}
    Attention(\bm{Q},\bm{K},\bm{V}) = \bm{\alpha} \bm{V}, \quad\bm{\alpha} = softmax(\frac{\bm{Q}\bm{K}^T}{\sqrt{d_k}}),
\end{split}
\end{equation}
where $\bm{\alpha} \in R^{N\times N}$ is the attention matrix.

Disregarding the Softmax function and the normalization operator, the output at index $i$ of the attention mechanism is
\begin{equation}
\label{equa:attenqkv}
\begin{split}
y_i &= \sum_{j=1}^{N}Q_iK_jV_j=\sum_{j=1}^{N}f_{t,q}(x_{q,i}) f_{t,k}(x_{v,j}) f_{t,v}(x_{v,j}),
\end{split}
\end{equation}
where $f_{t,q}$, $f_{t,k}$ and $f_{t,v}$ are the linear projections and $\bm{x}_q$ and $\bm{x}_v$ are input sequences.


\subsection{Mamba}
State space models (SSMs) map the input sequence $x_t \in R$ into $y_t \in R$ through a hidden state $h_t \in R^{D}$, which can be formulated as
\begin{equation}
\begin{split}
    h_t=\mathbf{\overline{A}}h_{t-1} + \mathbf{\overline{B}}x_{t},
\end{split}
\end{equation}
\begin{equation}
\begin{split}
    y_t=\mathbf{C}h_t,
\end{split}
\end{equation}
where $\mathbf{\overline{A}}$ and $\mathbf{\overline{B}}$ are the discretized parameter given the parameters $\mathbf{A}$, $\mathbf{B}$ and $\mathbf{\Delta}$:
\begin{equation}
\label{equa:abar}
\begin{split}
    \mathbf{\overline{A}} = exp(\mathbf{\Delta}\mathbf{A}),
\end{split}
\end{equation}
\begin{equation}
\label{equa:bbar}
\begin{split}
    \mathbf{\overline{B}} = (\mathbf{\Delta A})^{-1}(exp(\mathbf{\Delta A}) - \mathbf{I})\cdot\mathbf{\Delta B}.
\end{split}
\end{equation}

SSMs can also be calculated in the convolutional mode, which can be formulated as
\begin{equation}
\label{equa:conv}
\begin{split}
    \mathbf{\overline{K}}=(\mathbf{C}\mathbf{\overline{B}},\mathbf{C}\mathbf{\overline{A}}\mathbf{\overline{B}},\dots,\mathbf{C}\mathbf{\overline{A}}^k\mathbf{\overline{B}},\dots), \quad
    \bm{y}=\bm{x}*\mathbf{\overline{K}}.
\end{split}
\end{equation}

The system described above is based on a Linear Time Invariance (LTI) system. The recent work Mamba \cite{mamba} integrates a selective scan mechanism, deriving $\mathbf{B}$, $\mathbf{C}$ and $\mathbf{\Delta}$ from the input sequence, enabling the model to focus on different aspects of the input data. This calculation can be formulated as follows:
\begin{equation}
\label{equa:bi}
\begin{split}
    \mathbf{B_i} = S_B(x_i),
\end{split}
\end{equation}
\begin{equation}
\label{equa:ci}
\begin{split}
     \mathbf{C_i} = S_C(x_i), 
\end{split}
\end{equation}
\begin{equation}
\label{equa:di}
\begin{split}
    \mathbf{\Delta_i} = Softplus(S_{\Delta}(x_i)),
\end{split}
\end{equation}
where $S_B$, $S_C$ and $S_\Delta$ are linear projections and $i$ denotes the index of $i-th$ element of the sequence. Subsequently, the parameters of $\mathbf{\overline{A}_i}$, $\mathbf{\overline{B}_i}$ can be calculated following (\ref{equa:abar}) and (\ref{equa:bbar}). Mamba has been shown to offer performance comparable to Transformer while achieving significantly lower computational and memory complexity. 

\subsection{CrossMamba}
Since Mamba is specifically designed to capture long-term dependencies within a single sequence, it cannot be directly applied to target sound extraction tasks in the same manner as Transformers, which leverage the cross-attention mechanism to capture the dependencies between the clues and the audio mixture. To address this, we propose CrossMamba, following the analysis of the hidden attention mechanism of Mamba in \cite{mamba_atten}, which generates the hidden attention matrix by reformulating (\ref{equa:conv}) into the following matrix form:
\begin{equation}
    \bm{y}=\bm{\alpha}\bm{x},
\end{equation}
\begin{equation}
\label{equa:matrix}
\begin{split}
    \bm{\alpha} = \begin{bmatrix}
       \mathbf{C_1}\mathbf{\overline{B}_1} & 0 & \cdots & 0 \\
       \mathbf{C_2}\mathbf{\overline{A}_2}\mathbf{\overline{B}_1} &  \mathbf{C_2}\mathbf{\overline{B}_2} & \cdots & 0\\
       \vdots & \vdots & \ddots & 0\\
       \mathbf{C_N}\prod_{k=2}^{N}\mathbf{\overline{A}_k}\mathbf{\overline{B}_1} &  \mathbf{C_N}\prod_{k=3}^{N}\mathbf{\overline{A}_k}\mathbf{\overline{B}_2} & \cdots & \mathbf{C_N}\mathbf{\overline{B}_N}
    \end{bmatrix},
\end{split}
\end{equation}
where $N$ is the sequence length and $\bm{\alpha}$ is the hidden attention matrix in Mamba. 
The element of $\bm{\alpha}$ at row $i$ and column $j$ can be calculated as.
\begin{equation}
\label{equa:atten}
\alpha_{i,j}=\mathbf{C_i}\prod_{k=j+1}^{i}\mathbf{\overline{A}_k}\mathbf{\overline{B}_j}.
\end{equation}

By substituting (\ref{equa:abar}), (\ref{equa:bbar}), (\ref{equa:bi}), (\ref{equa:ci}) and (\ref{equa:di}) into (\ref{equa:atten}), the attention value can be computed as
\begin{equation}
\label{equa:atten_full}
\begin{split}
\alpha_{i,j}=&S_C(x_i)(Softplus(S_\Delta(x_j)) A)^{-1}\\&(exp(Softplus(S_\Delta(x_j))\mathbf{A}) - \mathbf{I}) \\ &(Softplus(S_\Delta(x_j))S_B(x_j))\\
&\prod_{k=j+1}^{i}exp(Softplus(S_{\Delta}(x_k))\mathbf{A}).
\end{split}
\end{equation}

We define that
\begin{equation}
\label{equa:qi}
\begin{split}
Q_i = S_C(x_i)=f_{m,q}(x_i), 
\end{split}
\end{equation}
\begin{equation}
\label{equa:ki}
\begin{split}
K_j &= (Softplus(S_\Delta(x_j)) A)^{-1}\\&(exp(Softplus(S_\Delta(x_j))\mathbf{A}) - \mathbf{I}) \\ &(Softplus(S_\Delta(x_j))S_B(x_j))\\
&=f_{m,k}(x_j),
\end{split}
\end{equation}
\begin{equation}
\label{equa:hi}
\begin{split}
H_{i,j} &= \prod_{k=j+1}^{i}exp(Softplus(S_{\Delta}(x_k))\mathbf{A}),
\end{split}
\end{equation}
\begin{equation}
\label{equa:vi}
\begin{split}
V_i = x_i, 
\end{split}
\end{equation}
then 
the output of Mamba can be represented as
\begin{equation}
\label{equa:mambaqkv}
\begin{split}
y_i = \sum_{j=1}^{i}\alpha_{i,j}x_j = \sum_{j=1}^{i}Q_iK_jH_{i,j}V_j.
\end{split}
\end{equation}

The cross-attention mechanism captures the dependencies of two sequences $\bm{x}_q$ and $\bm{x}_v$ by using $\bm{x}_q$ to generate the query and the $\bm{x}_v$ to calculate the key and value. By applying (\ref{equa:qi}) to $\bm{x}_q$, and substituting $\bm{x}_v$ into (\ref{equa:ki}), (\ref{equa:hi}) and (\ref{equa:vi}), CrossMamba, which implements the cross-attention mechanism, can be formulated as:
\begin{equation}
\label{equa:crossmambaqkv}
\begin{split}
y_i 
&=\sum_{j=1}^{i}f_{m,q}(x_{q,i}) f_{m,k}(x_{v,j}) H_{i,j}x_{v,j}.
\end{split}
\end{equation}

The difference between CrossMamba and the original Mamba is that in the original Mamba, $S_C$, $S_B$ and $S_{\Delta}$ are all applied to a single input sequence, while CrossMamba specifically applies the linear projection $S_C$ to the sequence intended to be the query in the cross-attention mechanism.

It can be seen that (\ref{equa:mambaqkv}) is a causal version of (\ref{equa:attenqkv}), which is often referred to as the masked attention, with the distinction that in (\ref{equa:mambaqkv}) $H_{i,j}$ controls the significance of the recent $i-j$ elements in the input sequence. We define this causal formulation as:
\begin{equation}
\label{equa:crossmamba}
\begin{split}
\bm{y} = CrossMamba(\bm{x}_q, \bm{x}_v).
\end{split}
\end{equation}
For the calculation of non-causal cross-attention, we employ the bi-directional form of (\ref{equa:crossmamba}). The non-causal cross-attention is then derived by summing the forward and backward forms, which can be expressed as:
\begin{equation}
\label{equa:nccrossmambaqkv}
\begin{split}
\bm{y} &=Bi\textit{-}CrossMamba(\bm{x}_q, \bm{x}_v)\\ 
&= CrossMamba(\bm{x}_q, \bm{x}_v) + \\
& flip(CrossMamba(flip(\bm{x}_q), flip(\bm{x}_v))).
\end{split}
\end{equation}

\begin{figure}[t]
\begin{minipage}[b]{0.9\linewidth}
  \centering
  \centerline{\includegraphics[width=5.3cm]{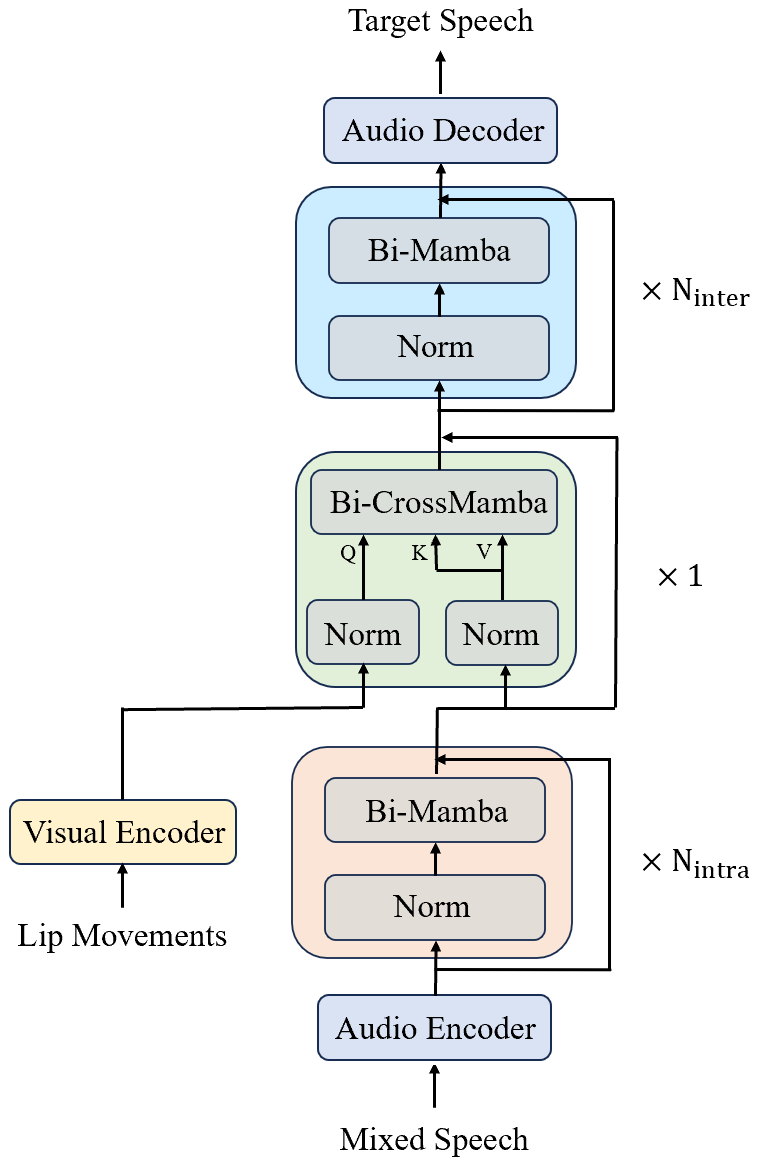}}
\end{minipage}
\caption{The structure of CrossMamba based AV-SepFormer.}
\label{fig:avmamba}
\end{figure}
\section{CrossMamba for Target Sound Extraction}
\label{CMTSE} 
The proposed CrossMamba\footnote{https://github.com/WuDH2000/CrossMamba} is implemented in two representative target sound extraction methods: AV-SepFormer, which leverages lip embeddings to extract the target sound from audio mixtures, and Waveformer, a real-time method that utilizes sound class labels for target sound extraction.
\subsection{CrossMamba for AV-SepFormer}
In AV-SepFormer, 1D convolutional layers are utilized to extract audio features, which are subsequently divided into chunks. A pre-trained visual encoder is employed to extract visual features. In the Separator, the audio features first pass through $N_{intra}$ IntraTransformer layers and are then fused with the visual features in a CrossModalTransformer layer. Subsequently, $N_{inter}$ InterTransformer layers are applied to capture inter-chunk information. More detailed implementation information can be found in \cite{avsepformer}. 

We use the bi-directional Mamba blocks from \cite{mamba}, which include a bi-directional state space model and RMSNorm, as replacements for the IntraTransformer and InterTransformer layers. The bi-CrossMamba block, which incorporates the bi-directional cross-attention-based state space model proposed above and RMSNorm, serves as a replacement for the CrossModalTransformer layer. The structure of CrossMamba based AV-Sepformer is illustrated in Fig \ref{fig:avmamba}.

\subsection{CrossMamba for Waveformer}
Waveformer employs 1D convolutional layers and 10 dilated causal convolution (DCC) layers \cite{dcc} to encode the input audio into features. Subsequently, a Transformer decoder layer fuses the embedding of the sound class label with the audio feature. Finally, deconvolution layers convert the audio chunks back into the time domain. Implementation details can be found in \cite{waveformer}.

\begin{figure}[t]
\begin{minipage}[b]{0.9\linewidth}
  \centering
  \centerline{\includegraphics[width=4.2cm]{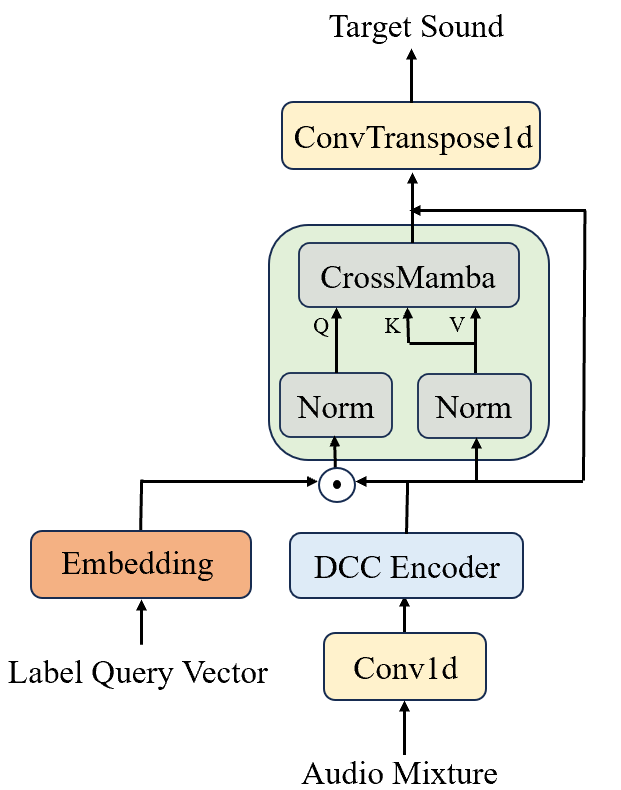}}
\end{minipage}
\caption{The structure of CrossMamba based Waveformer.}
\label{fig:avwaveformer}
\end{figure}

Since Waveformer is a real-time target sound extraction method that requires the model to be causal, we replace the Transformer decoder layer with the Causal CrossMamba block, which consists of the causal cross-attention-based state space model proposed above and RMSNorm. The structure of CrossMamba based Waveformer is shown in Fig \ref{fig:avwaveformer}.



\section{Experiments}
\label{experiments}
The separation performance of CrossMamba based AV-SepFormer and Waveformer is evaluated under their original experiment setups. Additionally, we compare the model size and the Multiply–Accumulate Operations (MACs) to show the computational efficiency of CrossMamba.
\subsection{CrossMamba for AV-SepFormer}
\subsubsection{Implementation setup}
The CrossMamba based AV-SepFormer is compared with the original AV-SepFormer at two scales: a large scale with $d_{model}$ equaling to $256$ and a small scale with $d_{model}=128$. We also establish two different scales of CrossMamba with the same $d_{model}$. The CrossMamba based AV-SepFormer is referred to AV-SepMamba.

The experimental setup follows that of AV-SepFormer, with models trained and tested on the VoxCeleb2 dataset \cite{voxceleb2}. The loss function used is the negative scale-invariant signal-to-noise ratio (SI-SNR). SI-SNR is also employed as the evaluation metric. More detailed information on the experimental setup can be found in \cite{avsepformer}.

\subsubsection{Resuls}
Table \ref{tab:sisnr-avsepformer} demonstrates the performance of different scales of AV-SepFormer and AV-SepMamba. Although the model size of AV-SepMamba-large is slightly higher than that of AV-SepFormer-large, AV-SepMamba-large has 60\% fewer MACs. This reduction is due to CrossMamba and Mamba's linear inference complexity, which is lower than the quadratic complexity of the attention mechanism. Besides, AV-SepMamba-large achieves a higher SI-SNR. For the smaller models, AV-SepMamba-small has a 32\% smaller model size and 73\% fewer MACs compared to AV-SepFormer-small, while also achieving a higher SI-SNR. Table \ref{tab:sisnr-avsepformer} demonstrates that CrossMamba based AV-SepFormer can achieve comparable or even better SI-SNR than Transformer-based models with much lower resource costs.



\subsection{CrossMamba for Waveformer}
\subsubsection{Implementation setup}
The Implementation setup follows that of \cite{waveformer}. The target sounds are sourced from the FSD Kaggle 2018 dataset \cite{fsd}, while the background noise comes from the TAU Urban Acoustic Scenes 2019 dataset \cite{tau}. The loss function is composed of 90\% negative SNR and 10\% negative SI-SNR, with the evaluation metric being the SI-SNR improvement over the signal mixture (SI-SNRi). We compare the CrossMamba based Waveformer models with the original Waveformer models under two configurations: a larger setup with an encoder dimension $E=512$ and a decoder dimension $D=256$, and a smaller setup with $E=512$ and $D=128$. The CrossMamba based Waveformer is refered to as WaveMamba for simplicity.

\subsubsection{Resuls}
\begin{table}[t]
\caption{SI-SNR of different scale of AV-SepFormer based on Transformers and CrossMambas}
\renewcommand\arraystretch{1.2}
\begin{center}
\begin{tabular}{c c c c}
\toprule
\textbf{Method}& SI-SNR (dB) & Params (M) & MACs (G/s)\\
\midrule
AV-SepFormer-large & 13.04 & \textbf{29.63} & 414.08\\
AV-SepMamba-large & \textbf{13.20} & 30.36 & \textbf{165.95} \\
\midrule
AV-SepFormer-small & 12.11 & 13.32 & 172.09\\
AV-SepMamba-small & \textbf{12.21} & \textbf{9.08} & \textbf{45.88}\\
\bottomrule
\end{tabular}
\label{tab:sisnr-avsepformer}
\end{center}
\end{table}
\begin{table}[t]
  \caption{SI-SNRi of different scale of Waveformer based on Transformers and CrossMambas on single target extraction}
  \renewcommand\arraystretch{1.2}
  \tabcolsep=0.11cm
   \centering
  \begin{tabular}{>{\centering\arraybackslash}p{3cm} >{\centering\arraybackslash}p{1.6cm} >{\centering\arraybackslash}p{1.5cm} >{\centering\arraybackslash}p{1.5cm}}
    \toprule
    \textbf{Method} & SI-SNRi (dB) & Params (M) & MACs (G/s)
                            \\
    \midrule
Waveformer-large & 9.43 & 3.88 & 15.80 \\
WaveMamba-large & \textbf{9.54} & \textbf{3.66} & \textbf{12.74}\\
\midrule
Waveformer-small & 9.26 & 3.29 & 12.54\\
WaveMamba-small & \textbf{9.67} & \textbf{3.24} & \textbf{11.81}\\
\bottomrule
  \end{tabular}
\label{tab:sisnr-waveformer}
\end{table}
Table \ref{tab:sisnr-waveformer} presents the SI-SNRi values on the single target extraction task. Since Waveformer is designed as a lightweight, real-time model with a small size and efficient computation, the reduction in model size and MACs with CrossMamba is not significant. Nonetheless, CrossMamba-based models achieve higher SI-SNRi with fewer model parameters and MACs compared to Waveformer models with equivalent encoder and decoder dimensions.

\section{Conclusion}
\label{conclusion}
In this paper, we propose CrossMamba, which incorporates the cross-attention mechanism to capture dependencies between two sequences. This enables the replacement of Transformers for feature fusion in target sound extraction models, offering higher computational and memory efficiency. We follow the analysis of the hidden attention mechanism in Mamba, divide the Mamba formulation into the query, key and value and generate the query from the clue and the key and value from the audio mixture. Experimental results on two representative target sound extraction methods
demonstrate that CrossMamba achieves better performance with a reduced computational load. Furthermore, the design of CrossMamba is based on the principles of the cross-attention mechanism rather than the specifics of target sound extraction tasks, indicating its potential applicability to a wide range of other cross-attention-based tasks in the future.
\bibliographystyle{IEEEtran}
\bibliography{refs.bib}
\end{document}